\documentclass[prl,aps,showpacs,superscriptaddress,twocolumn]{revtex4}
\usepackage{color} 
\usepackage{times}
\usepackage{amssymb} 
\usepackage{amsmath} 
\usepackage{ragged2e}
\usepackage{graphicx}
\usepackage{epstopdf}
\usepackage[english]{babel}
\usepackage{mathrsfs}
\usepackage{textcomp}
\usepackage{amsthm}
\usepackage{amsfonts}
\usepackage{soul}
\usepackage{braket} 

\newcommand{\be}{\begin{equation}}
\newcommand{\ee}{\end{equation}}
\newcommand{\bea}{\begin{eqnarray}}
\newcommand{\eea}{\end{eqnarray}}
\newcommand{\ba}{\begin{array}}
\newcommand{\ea}{\end{array}}

\newcommand{\av}[1]{\left\langle #1 \right\rangle}
\newcommand{\ct}{\dagger}

\begin{document}

\title{Clustered Superfluids in the One Dimensional Bose-Hubbard model with extended correlated hopping}

\author{Julia Stasi\'nska}
\affiliation{Institute of Physics, Polish Academy of Sciences, Aleja Lotnik\'ow 32/46, 02-668 Warsaw, Poland}\email{julsta@gmail.com}

\author{Omjyoti Dutta}
\affiliation{Donostia International Physics Center DIPC, Paseo Manuel de Lardizabal 4, 20018 Donostia-San Sebastian, Spain}

 \author{Luca Barbiero}
\affiliation{ICFO - Institut de Ciencies Fotoniques, The Barcelona Institute of Science and Technology, Av. Carl Friedrich Gauss 3, 08860 Castelldefels (Barcelona), Spain}

\author{Maciej Lewenstein}
\affiliation{ICFO - Institut de Ciencies Fotoniques, The Barcelona Institute of Science and Technology, Av. Carl Friedrich Gauss 3, 08860 Castelldefels (Barcelona), Spain}
\affiliation{ICREA, Pg. Llu\'is Companys 23, 08010 Barcelona, Spain}

\author{ Ravindra W. Chhajlany} 
\affiliation{ Faculty of Physics, Adam Mickiewicz University, Umultowska 85, 61-614 Pozna{\'n}, Poland}

\date{\today}

\begin{abstract}
Bosonic lattice systems with non-trivial interactions represent an intriguing platform to study exotic phases of matter. Here, we study the effects of extended correlated hopping processes  in a system of bosons trapped in a lattice geometry. The interplay between single particle tunneling terms, correlated hopping processes and on-site repulsion is studied by means of a combination of exact diagonalization, strong coupling expansion and cluster mean field theory. We identify a rich ground state phase diagram where, apart the usual Mott and superfluid states, superfluid phases with interesting clustering properties occur. 
 \end{abstract}

\maketitle

\section{Introduction}\label{sec_introduction}

Clustering in gapless ultracold bosonic systems has recently garnered intensive interest thanks, on one hand, to the experimental realization of quantum liquids composed of droplets both in dipolar systems \cite{pfau, ferlaino1} and bosonic mixtures \cite{leticia,fattori}, and on the other hand to the very recent realization of supersolid states \cite{tanzi,chomaz,pfau1}. As it has also been determined theoretically, the achievement of such a state of matter requires  specific tuning of the interactions~\cite{petrov, lima}. These features have remarkably also been identified in a locally interacting bosonic mixture trapped  in  a purely one dimensional lattice  \cite{bruno1,bruno}. This kind of systems is usually and properly described by a the Bose-Hubbard model  (BHM) which, for a single bosonic species, represents the simplest Hamiltonian  where the interplay between interaction and  kinetic energy of particles gives rise to fundamental quantum effects. Indeed, as an early motivation, the BHM gained attention as a means to   understand the effects of repulsive interactions on a superfluid phase \cite{Knollmann1963} while later on it has been both theoretically \cite{fisher} and experimentally \cite{greiner} demonstrated, that for integer density values the competition between interaction and kinetic energy allows for a pure quantum phase transition between a Mott insulator state and a superfluid. Furthermore this model has turned out to be relevant  to the description of Josephson-junction arrays~\cite{bruder2005}, quantum magnets~\cite{giamarchi2007} and photonic systems \cite{shahaf}. Renewed theoretical interest has been driven by highly tunable experimental realizations of close-to-ideal generalized BHMs in systems of ultra-cold atoms trapped in optical lattices \cite{lewenstein_rev}, where other terms such as long range interactions \cite{ferlaino}, or higher order hopping processes \cite{nagerl,Naegerl} are important (apart from the usually dominant on-site interactions and hopping processes) and lead to richer phase diagrams \cite{Dutta-review}. \\
\begin{figure}[t]
\includegraphics[width=0.99\columnwidth]{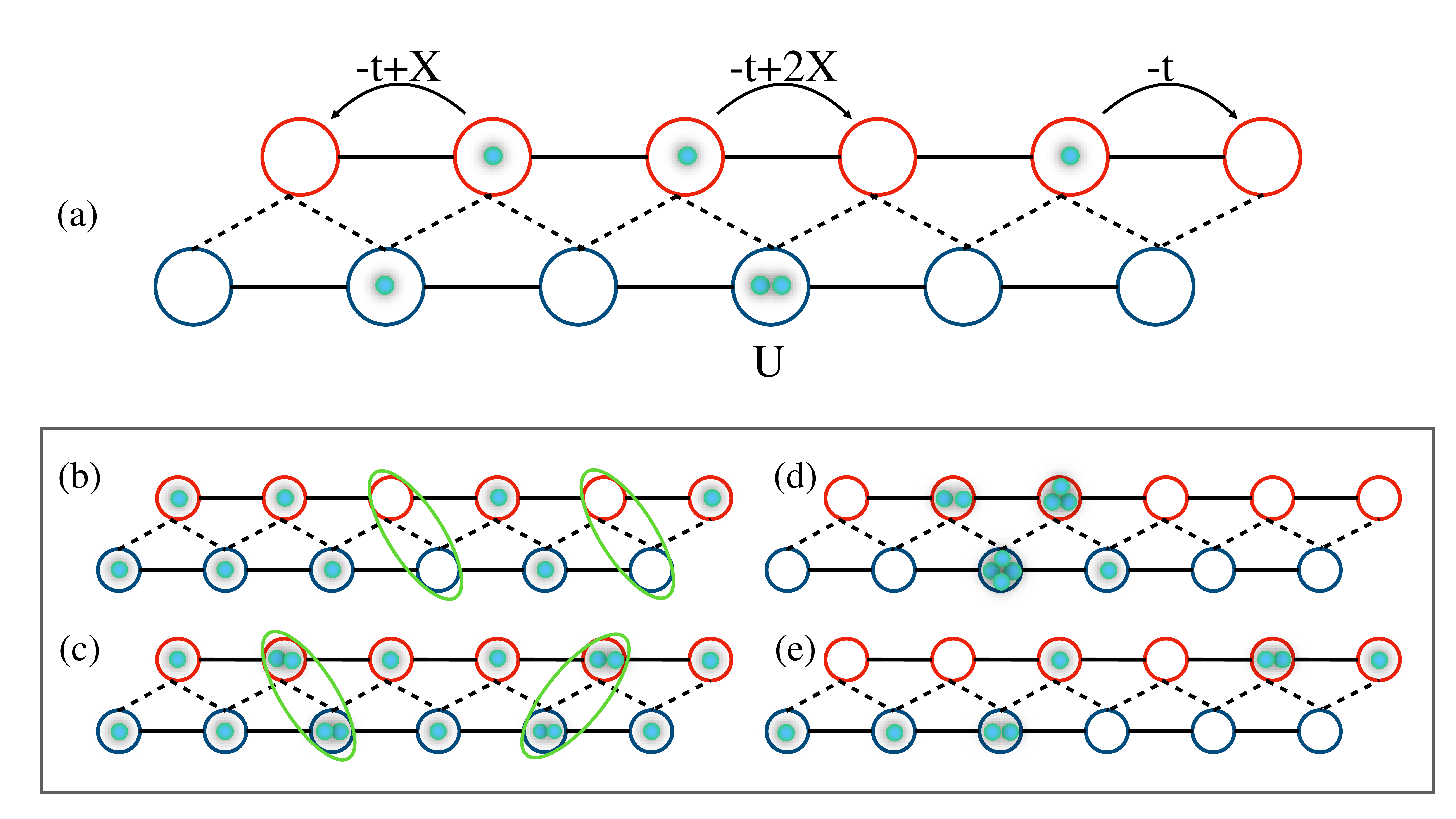} 
\caption{(Color online) Schematic of the lattice consisting of bosons residing on a 1-dimensional triangular lattice described by the interacting Hamiltonian (\ref{eq_model}). (a) Shown is a particular configuration of sites filled with particles (green) and the various elementary processes in the system. The bottom frame shows exemplary configurations corresponding to the non-standard phases of the model. (b) Paired-hole superfluid below unit filling: every particle on the lower layer (blue) is always followed by a particle on the upper lattice (red) as a result of which pairs of coherently moving bound nearest neighbor holes characterize the system.(c) Paired-particle superfluid: above unit filling, nearest neighbour doubly-occupied sites are paired and move coherently through the lattice. (d) Unbounded droplet phase:  most of the particles form a cluster of occupied consecutive sites. (e) Phase separation:  particles form two separated regions in space, one where only the lower layer (blue) is occupied and the other where the upper layer (blue) is occupied.}  
\label{fig:model}
\end{figure}
In this context, the correlated hopping (CH) terms, describing how the single particle tunneling amplitude is influenced by the position of another particle, represent a very intriguing kind of interaction. On one hand, their effect was investigated mainly in fermionic systems to get insights initially on high-temperature hole-superconductors~\cite{Micnas,Hirsch}.  It has been further shown that in one dimensional  fermionic systems CHs can give rise to exotic metallic \cite{Grzybowski2012}, superconductor \cite{aligia,anfossi1,anfossi2} and magnetic \cite{anfossi3, anfossi4} regimes as well as to symmetry protected topological states \cite{barbiero, fazzini} and lattice gauge theories \cite{gorg}. On the other hand, the investigation of bosonic systems in presence of such kind of interaction started just recently. Nonetheless, it has already been shown that CHs may lead to the formation of stable condensates of boson pairs~\cite{Mila,Xu,Eckholt} without direct attractive interactions, paired superfluids \cite{Sowiiski, Santos}, supersolid regimes \cite{maik}, scar states \cite{hudomal}, and gauge constrained confined phases \cite{santos1,barbiero1, schweitz}.\\
A special kind of CHs are the ones, called extended correlated hopping (ECH) terms, where the tunneling range exceeds the usual nearest neighbor approximation. It is worth mentioning that such interaction terms naturally appear in rigorous descriptions of long-range \cite{barbiero2,li} kinetically constrained one dimensional systems, strongly interacting Raman coupled bosonic mixtures \cite{Bilitewski2016} and semi-synthetic zigzag optical lattices \cite{Anisimovas2016}. Furthermore a recent paper~\cite{Chhajlany_fermions} has introduced a one-dimensional fermionic lattice construction for which contact interaction actually leads to an effective three site extended correlated hopping thus realizing a variant of the Bariev model \cite{Bariev}. It was shown that in such a model a rich phase diagram emerges when ECHs destructively interfere with single-particle hopping. On the other hand, bosonic systems with extended correlated hoppings have been treated only in the hard-core regime both in square \cite{Mila} and triangular \cite{Xu} lattices.\\
For this reason in this article, we consider the effects of extended correlated hopping on a one dimensional bosonic system with loosened on-site constraints, {\it i.e.} soft-core bosons, instead. We start our analysis by describing the model and its limiting cases where the ground state quantum regime is known. This helps us to perform a first numerical analysis based on fidelity behaviour determined via exact diagonalization calculations. In addition to this, we perform a strong coupling expansion to get further insight on the strongly repulsive regime. Finally we employ the cluster mean-field method as a complementary tool to enforce our results. Thanks to this approach we are able to derive a large variety of phases appearing in the quantum phase diagram as a function of the interplay among single particle tunneling, extended correlated hopping and on-site interaction terms. In particular, apart from the usual Mott and superfluid phase, we discover the presence of a paired-holes superfluid, a paired-particles superfluid and a superfluid of unbounded droplets -- all of which are characterized by  non-trivial clusterization properties.
\subsection{The model}\label{subsec_themodel}
We consider the following one-dimensional lattice Hamiltonian:
\begin{eqnarray}\label{eq_model}
H & = & \sum_{i}^{L} \big[-t(b_i^{\ct}b_{i+2} + h.c.) + X (b_i^{\ct}n_{i+1} b_{i+2}+h.c.).\nonumber\\
& + &\frac{U}{2} n_i (n_{i}-1)+\mu n_i\big]
\end{eqnarray}
where $b_i^{\dagger}/b_i$ is the creation/annihilation operator of a boson in site $i$ of a lattice composed by L sites. The first term in the Hamiltonian describes direct single particle hopping between next nearest neighbours sites with amplitude $t$. On the other hand $X$ is the amplitude of the extended correlated hopping, thus capturing the processes where the next nearest neighbours hopping is conditioned by a finite occupation between the two sites involved in the particle tunneling. The remaining terms describe the onsite repulsion $U$ and chemical potential $\mu$. The absence of nearest-neighbour tunneling preclude the mixing between particles in even and odd sites. Therefore, the model can be viewed as a triangular bosonic ladder with leg-conserving number of particles. In our work we consider the even/odd sublattices to be of equal lengths $L/2$, and the total number of particles in the system to be $N=N_{\rm{even}}+N_{\rm{odd}}$ thus fixing the system density $\rho=N/L$. Note that we do not \textit{a priori} presume equal occupation of both sublattices.\\We further recall here that the fermionic model \cite{Chhajlany_fermions} is analytically diagonalizable when $X/t=1$ and $U=0$ for arbitrary filling. For this parameters choice, a fermion on one sublattice cannot hop over a nearest neighbor fermion on the other sublattice, i.e. the matrix element $-t+Xn_{i+1}=0$ for $n_{i+1}=1$. This strong additional symmetry constraint facilitates the analytical solution. For the same parameters, this symmetry is generically broken for a bosonic system since sites can be populated by more than one particle. 
\begin{figure}
\includegraphics[width=0.95\columnwidth]{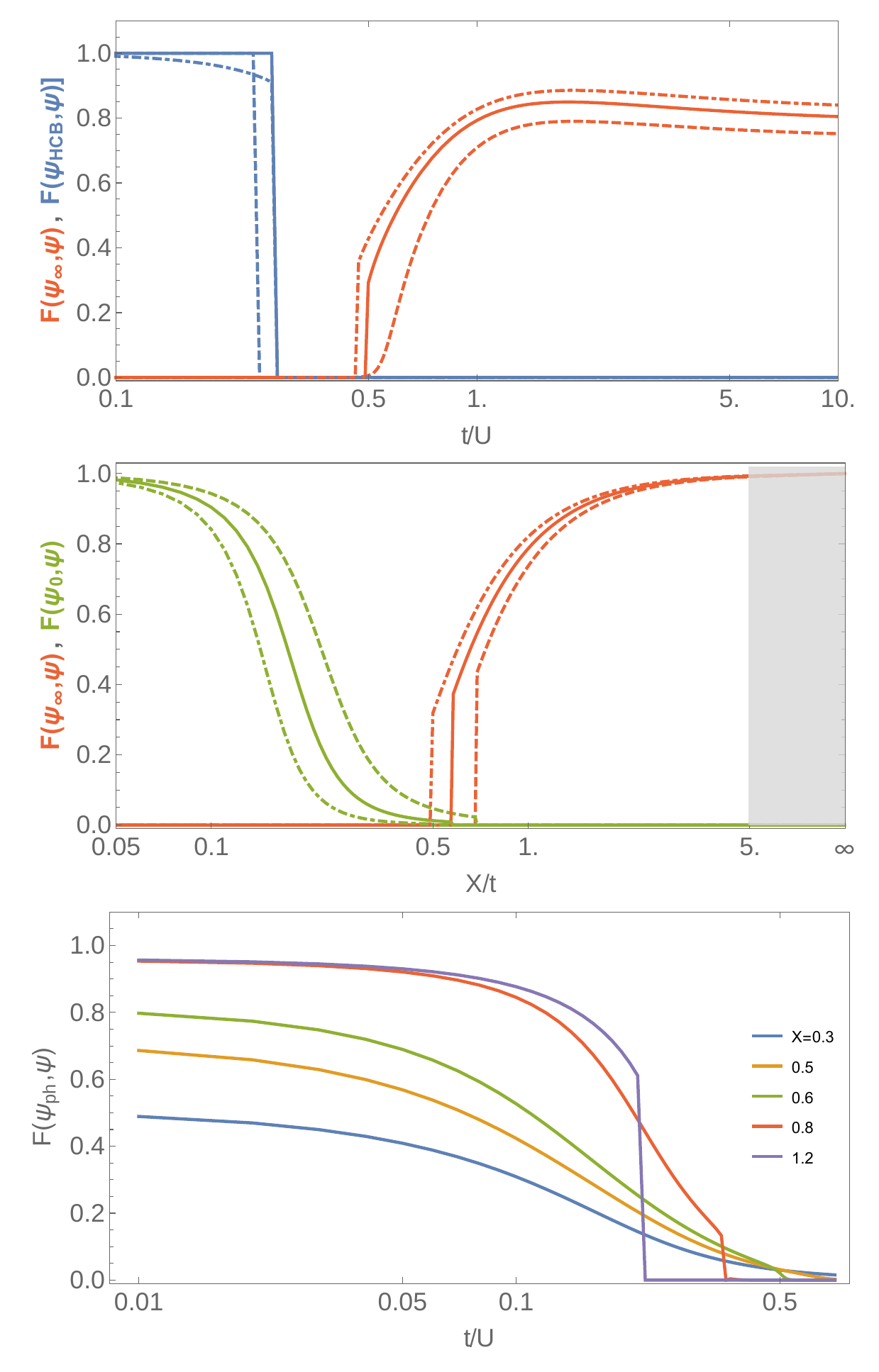}
\caption{\textit{top panel}: fidelity of the ground state with the state in the hard-core boson limit (HCB) $|\psi_{\rm HCB} \rangle$ (blue) and droplet superfluid state $|\psi_{\infty}\rangle$ (orange) as a function of $t/U$ for $X=t$ for unit  filling (continuous),  filling $(N-1)/L$ (dashed), and filling $(N+1)/L$ (dotted). The state $|\psi_{\rm HCB} \rangle$ corresponds to Mott insulator $|\psi_{\rm Mott} \rangle$ for $\rho=1$, $|\psi_{\rm ph} \rangle$ for $\rho<1$ and $|\psi_{\rm pp} \rangle$ for $\rho>1$. \textit{central panel}: fidelity with the standard superfluid phase $|\psi_0\rangle$ (green) and the droplet superfluid state $|\psi_{\infty}\rangle$ (orange) as a function of $t/U$ for $X=t$ for unit  filling (continuous),  filling $(N-1)/L$ (dashed), and filling $(N+1)/L$ (dotted). \textit{lower panel}: fidelity of the ground state with $|\psi_0\rangle$, $|\psi_{\rm ph}\rangle$ for $N/L=5/6$ and different values of $X/t$ as a function of $t/U$. All the results have been obtained via exact diagonalization calculation of systems with $L=12$}
\label{fig_fid}
\end{figure}
Nevertheless we note that there are known and intuitive solutions for some limiting cases of the model Eq.~(\ref{eq_model}). In analogy with standard Bose-Hubbard model, at unit filling and for strong on-site repulsion $U \gg t ,X$ the system is a Mott insulator (MI) $\ket{\psi_{\rm Mott}}$.  At the same time in absence of correlated hopping $X=0$ and for non-integer filling the system is always a standard superfluid (SF) phase $|\psi_0\rangle$ and it remains SF even for integer filling but low $U/t$. Furthermore for $X=t$, $N/L<1$ and hard core repulsion $t/U \to 0$ the particles behave effectively as fermions and hence enter in the so-called paired-hole superfluid (PHSF) state discussed in Ref.~\cite{Chhajlany_fermions}. In this state $|\psi_{\rm ph}\rangle$, any consecutive pair of bosons on the lattice is separated by an even number of empty (hole) sites. For such values of $X$, the motion of particles allowed by the Hamiltonian is constrained so that effectively pairs of nearest neighbor holes move whenever an allowed single hopping event  of a particle occurs. For small $U/t$, $N/L>1$ and again $X=t$ the paired-particles superfluid (PPSF) phase $|\psi_{\rm pp}\rangle$ appears, this latter being the analogous to the paired-holes superfluid for fillings larger than one. Finally a superfluid with unbounded droplets (DSF) $|\psi_{\infty}\rangle$ may occur as the ground state of the Hamiltonian (\ref{eq_model}) in the limit $t=0, U=0, X>0$. This DSF is characterized by a large spatial clustering of particles and a major contribution to the ground state coming from high-density clusters.


\section{Results}\label{Phase diagram}
\subsection{Exact diagonalization}\label{sub_diag}
To gain exact insight into the ground state properties of the many body Hamiltonian Eq.~(\ref{eq_model}), we first present an analysis based on exact diagonalization calculations. We consider the system at exactly unit filling $N=L$  as well as under the simplest deviations from this density, \textit{i.e.} $N=L\pm1$. The local bosonic Hilbert space is not truncated, {\it i.e.} we choose the maximal possible on site occupation to be $N$ and we impose periodic boundary conditions.\\
Thanks to known solutions, an immediate way to get insights about the phase diagram is achieved by computing the fidelitiy $F(\psi,\psi_{\rm ref}) = |\langle \psi_{\rm ref}|\psi\rangle|^2$ of the calculated ground state $|\psi\rangle$ with the various numerically extracted reference states $|\psi_{\rm ref}\rangle$ ( $|\psi_{\rm Mott} \rangle$, $|\psi_0\rangle$, $|\psi_{\rm ph}\rangle$, $|\psi_{\rm pp}\rangle$ and $|\psi_{\infty}\rangle$. As examples, in Fig. \ref{fig_fid} we show that this strategy allows us to properly capture the relevant effect induced by the ECH. Indeed it is possible to notice that, in the limit of strong on-site interaction, the Mott insulator known from the standard Bose-Hubbard model is additionally stabilized by $X$, which effectively decreases the hopping rate. This is especially visible close to $X=t$, for which the insulating phase turns out to extend up to $t/U \approx 0.3$ while only up to $t/U \approx 0.2$ for $X=0$. In addition to this we notice that the presence of ECHs allows for the occurring of different kind of gapless superfluid states. As visible in the central panel of Fig. \ref{fig_fid}, when $U=0$ and $X$ is weak the usual superfluid state is the dominant order, on the other hand larger $X/t$ ratios favor the presence of a droplet superfluid $|\psi_{\infty}\rangle$ which, as we will discuss more in details later, posses intriguing clustering properties. Moreover, for the case of non-integer densities, the same approach based on fidelity measurements allows to properly capture the extension of the paired-hole superfluid state. Indeed, as visible in Fig. \ref{fig_fid}, we are able to show that the PHSF survives well beyond the hard core $t/U\rightarrow 0$ limit even for relatively small $X/t$.
\begin{figure*}
\includegraphics[width=0.9\textwidth, trim=0ex 6ex 0ex 0ex, clip]{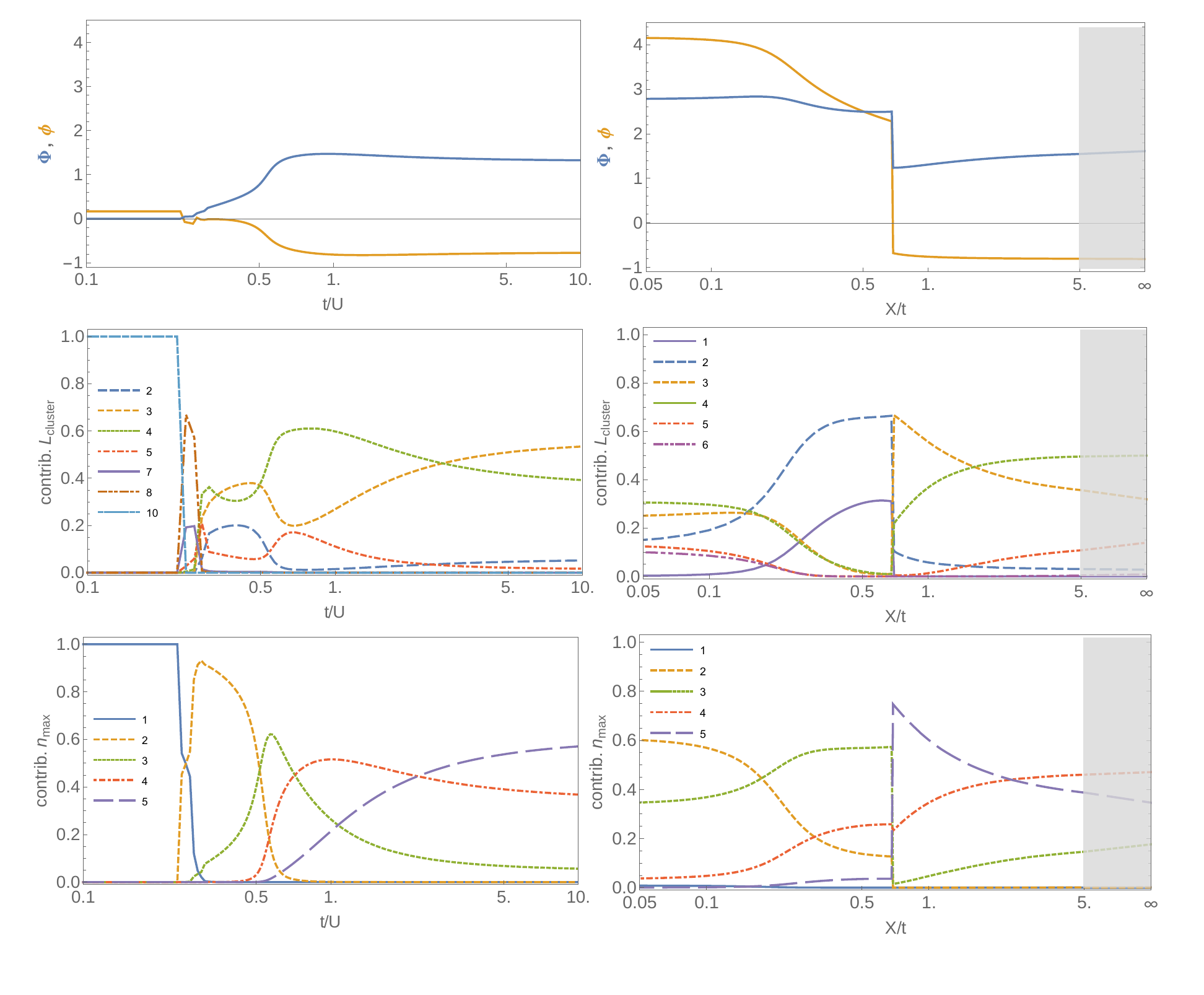}\\
\caption{\textit{first line}: One- and two-particle correlation function $\phi$ and $\Phi$, \textit{second line}: contribution from basis vectors with maximal cluster of length $L_{\rm cluster}$, \textit{third line}: contribution from basis vectors with maximal on-site occupation $n_{\rm max}$ as a function of $t/U$ at $X=t$(first column) and as a function of $X/t$ at $U=0$ (second column). All the results refer to systems with $L=12$ and $N=10$}
\label{fig4} 
\end{figure*}\\
In order to better characterize the phase diagram in the intermediate regimes, where the fidelity approach is supposed to be less accurate, in Fig. \ref{fig4} we also calculate the off-diagonal part of the one- $\phi=\sum_{j\neq i} \langle b^\dagger_i b_j\rangle$ and of the two-body $\Phi=\sum_{j\neq i} \langle b^\dagger_i b^\dagger_i b_jb_j\rangle$ density matrix. The latter quantities result to be very useful to distinguish between possible paired states. This turns out to be of crucial relevance when dealing with ground state clustering of bosons. In order to be even more accurate in characterizing these states, we further define the contribution (weight) of configurations with certain maximal on-site occupations $n_{max}$ and for which the maximal sequence of occupied sites (called maximal cluster) has a specific length $L_{cluster}$. For illustration consider the following vector in occupation number representation $\ket{1\,1\,0\,0\,1\,0\,1\,3\,1\,1\,1\,0}$. The maximal on-site occupation is $n_{max}=3$ while the maximal cluster is $[1\,3\,1\,1\,1]$ having a length $L_{cluster}=5$. In Fig. \ref{fig4} we combine the two aforementioned quantities to enforce our characterization of the phase diagram in the regime of density slightly smaller than one. For $X=t$, left column, the first thing to underline is that, while our results confirm the presence of a hole-paired superfluid for strong repulsion and intermediate ECHs, they also allow to capture that PHSF is characterized by almost null one- and two-particle correlation functions, and maximal contribution to the wave function from states characterized by cluster length $L_{\rm cluster}=N$ and maximal occupation $n_{\rm max}=1$. In the interval $0.25 \lesssim t/U \lesssim 0.5 $ we observe a change of sign of the one-particle correlation function $\phi$ as well as a small growth of two-particle correlation function $\Phi$. The latter is associated with the appearance of bosonic pairs, which is also confirmed by the fact that the maximal contribution to the wave function comes from states with maximal occupation $2$ or $3$. Yet, the pair character of the superfluid in this region is rather weak. Moreover, at $t/U \approx 0.31$ we observe a significant change in the structure of states contributing to the wave function: the states with long clusters no longer dominate, instead the states of cluster lengths $3$ or $4$ account for about $70\%$ of the wave function. This can be associated with the appearance of a superfluid composed by unbounded droplets, however of different kind (fragmented and with lower onsite density) as the fidelity with $\ket{\psi_{\infty}}$ is zero. The true superfluid quantum droplet phase starts to appear at $t/U \approx 0.5$, where it is linked to a significant growth of the absolute value of the correlation functions $\phi$ and $\Phi$. We also notice the change in the structure of the wave function indicated by the growing contribution from states having clusters of length 4 and 5 (3, 4 in the limit of large t/U) and maximal onsite occupation 3, 4 or 5, e.g. states with high-density clusters of type $\ket{\ldots 0\,1\,5\,4\,0 \ldots}, \ket{\ldots 0\,1\,4\,4\,1\,0 \ldots}$. Moreover we notice that the overlap with the state $|\psi_{\infty}\rangle$ (Fig. \ref{fig_fid}) increases to reach $0.8$ around $t/U=1.$\\
Let us now explore the case of $t/U\to\infty$ and varied $X/t$. The properties of the system in this limit are summarized in right column of Fig. \ref{fig4}. 
At weak correlated hopping $X/t \lesssim 0.2$ the ground state of the system is the standard SF as confirmed by large value of the correlation functions and almost unit fidelity with $|\psi_0\rangle$. In the interval $0.2 \lesssim X/t \lesssim 0.67$ the one- and two-particle correlation functions decrease slightly and the fidelity with $|\psi_0\rangle$ drops to zero. At the same time the predominant contribution to the wave function comes from states with short one- and two-site clusters with maximal on-site occupation of 3 and 4. This is due to phase separation, {\it i.e.} the particles belonging to each sublattice concentrate in spatially disjoint regions of length $L/2$ in which they are effectively described by the standard Bose-Hubbard model: $H_{\rm BH}=\sum_{{\rm odd}\ i=1}^{L/2} \left[-t (b_i^{\dagger} b_{i+2}+h.c.)+\frac{U}{2} n_i (n_i-1)\right]$ (and similarly for even sites) and form a superfluid. In the bulk of each such region only even/odd sites are occupied leading to single-site clusters, however at their interface two consecutive even-odd sites can be occupied, hence the 2-site clusters are present. An exemplary state, which may contribute to the ground state is $|1 0 3 0 1 0 0 2 0 2 0 1\rangle$. For even larger values of the correlated hopping $X/t\gtrsim 0.67$ the one-particle correlation function becomes negative and the two-particle correlation function drops slightly. We also observe sudden clustering of particles in clusters of length 3 and 4 with maximal occupation being 5, 4 or 3, which signals the transition to the DSF. This is confirmed by the fast increase of fidelity with $|\psi_{\infty}\rangle$, which reaches $0.5$ around $X/t \approx 0.7$ and $0.8$ for $X/t\approx 1$.
\begin{figure*}
\includegraphics[width=0.9\textwidth, trim=0ex 6ex 0ex 0ex, clip]{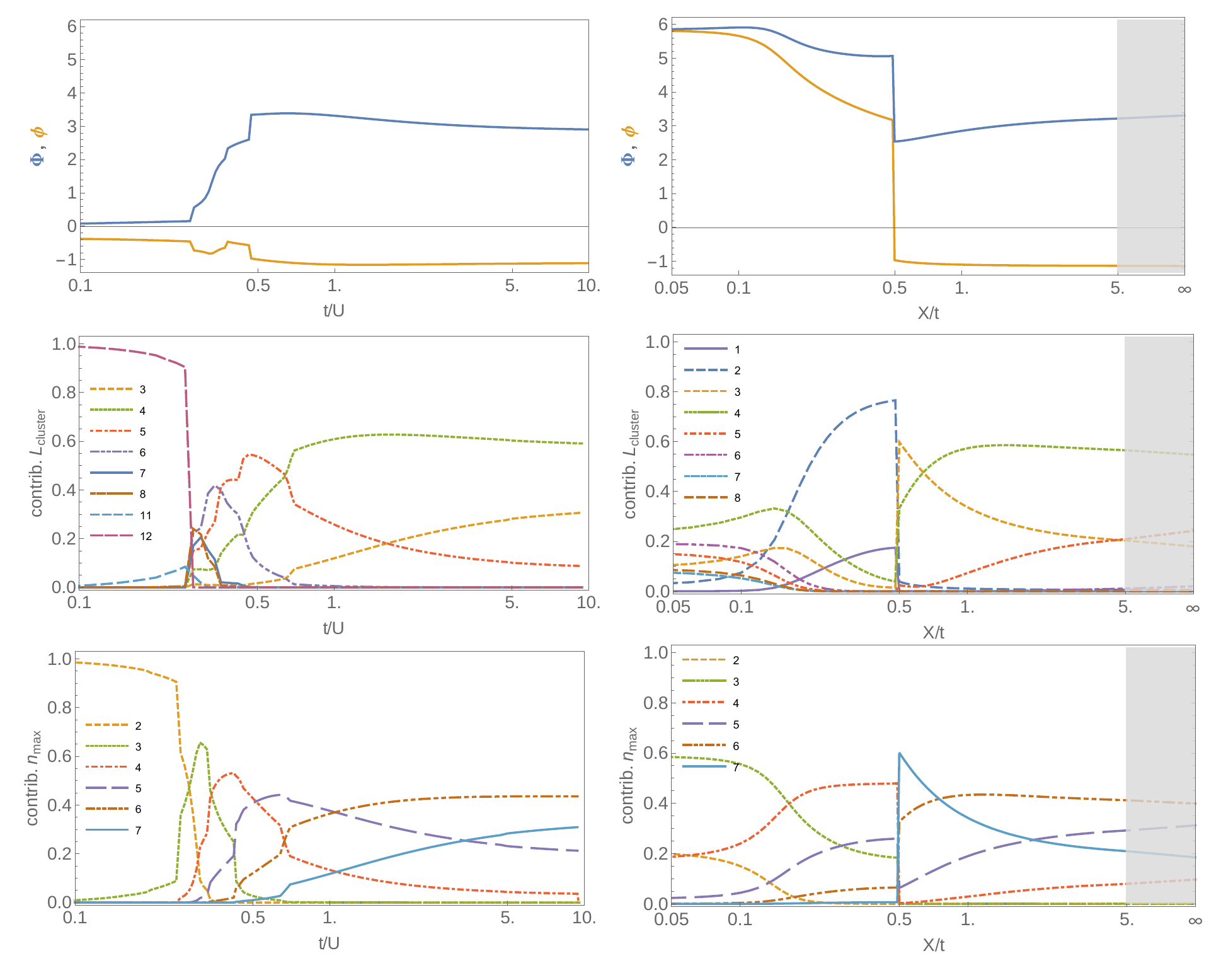}\\
\label{fig_edpropertiesL12n14U0}
\caption{\textit{first line}: One- and two-particle correlation function $\phi$ and $\Phi$, \textit{second line}: contribution from basis vectors with maximal cluster of length $L_{\rm cluster}$, \textit{third line}: contribution from basis vectors with maximal on-site occupation $n_{\rm max}$ as a function of $t/U$ at $X=t$(first column) and as a function of $X/t$ at $U=0$ (second column). All the results refer to systems with $L=12$ and $N=14$}
\label{fig_edpropertiesL12n14}
\end{figure*}\\
In Fig. \ref{fig_edpropertiesL12n14} we complement or analysis by studying the case of filling higher than one $N=L+2$. The most important difference with respect to the lower filling case is the nature of the ground state for $X=t$ and strong repulsion. While for $\rho<1$ the bosonic system behaves similarly to the fermionic case and its ground state is the PHSF, for $\rho>1$ we have a completely new situation. Exact diagonalization data show that the system with two additional particles above the unit filling becomes indeed a paired-particles superfluid, in which doubly-occupied sites are paired and particles belonging to such pair can only hop in a leapfrog manner, {\it i.e.} a particle hops over a doubly-occupied site as hopping over a single particle is suppressed by the correlated hopping term. The effective tunneling parameter associated with such hopping process is positive, {\it i.e.} $-t+2X=+t$, hence to minimize the energy the ground state has to be of the form $|\psi\rangle \propto \sum_i (-1)^i b_i^{\dagger}b_{i+1}^{\dagger} \ket{1\,1 \ldots}$. The structure of the wave function can thus be also inferred from the unit contribution of the states with cluster of length $L_{\rm cluster}=12$ and maximal occupation $n_{\rm max}=2$, and the negative sign of the one-particle correlation function. Note that close to $t/U\approx 0.1$ there appear contributions from states with clusters of length $11$ and maximal occupation $3$, {\it i.e.} states of the type: $|0\,2\,3\,1 \ldots\rangle$ in which the energy is lowered through formation of a low-density droplet.


\subsection{Strong coupling expansion}\label{sub_estimates}
\begin{figure*}
\includegraphics[width=1.05\textwidth]{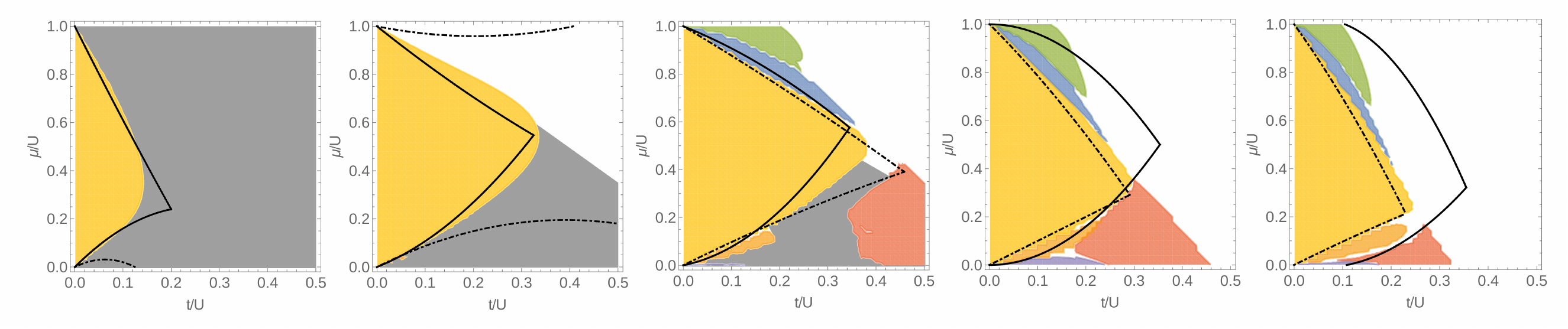}
\caption{Phase diagrams at different $X/t=0.0, 0.6, 0.8, 1.0, 1.2$ as a function of $\mu/U$ and $t/U$. The colors refer to the different phases obtained by combining the value of superfluid parameter $\phi_i=\langle b^{\dagger}_i\rangle +\sum_{j} \langle b_i^{\dagger}b_j\rangle $ on central sites and the properties of the wave function. In particular the colors are defined as follow: (red) phase separation, (orange) paired holes superfluid (PHSF) with two holes, (purple) paired holes superfluid (PHSF) with 4 holes, (yellow) Mott insulator (MI), (green) paired particles superfluid (PPSF) with 2 additional particles, (blue) paired particles superfluid (PPSF) with 4 additional particles, (gray) normal superfluid (SF). In white areas we were not able to unambiguously determine the properties of the system using the CM approach. Strong coupling expansion results are depicted in black: (dot-dashed) transition to PHSF or PPSF, (continuous) transition to standard superfluid.}
\label{PhaseDiags}
\end{figure*}

To estimate the extent of the phases identified in the previous section we perform the strong coupling expansion up to the second order in $t/U$ \cite{Freericks1996}. We rewrite the model Hamiltonian (\ref{eq_model}) as:
\begin{equation}
H = H_0-\frac{t}{U}\sum_{i=1}^{L} \left[b_i^{\ct}\left(1-\frac{X}{t} n_{i+1}\right)b_{i+2} + h.c. \right],
\label{model_sc}
 \end{equation}
where $\displaystyle H_0=\sum_{i=1}^L \left[\frac{1}{2} n_i(n_i-1) - \frac{\mu}{U} n_i\right]$ is diagonal in the number basis and the term proportional to $t/U$ is a perturbation.
We calculate the energies for the Mott insulator state of density $n_0=\rho$ 
\begin{equation}
\ket{\psi_{\mathrm{Mott}}(n_0)}=\ket{n_0 n_0 \ldots},
\label{Mott_psi}
\end{equation}
obtaining up to the second order in $t/U$:
\begin{eqnarray}\label{Mott_E}
E_{\mathrm{Mott}}&=&L \left[\frac{1}{2} n_0 (n_0-1)-\frac{\mu}{U} n_0\right]\\
&-&\frac{2t^2}{U} L n_0(1+n_0) \left(1-\frac{X}{t} n_0\right)^2.\nonumber
\end{eqnarray}
In the standard superfluid-Mott insulator transition, the variation from the Mott state leading to superfluidity is an addition of one hole/particle, however in the present case we assume the symmetric occupation of the two sublattices, hence the superfluid ansatz of the form:
\begin{eqnarray}
\ket{\psi_{\mathrm{Holes}}}&=&\frac{2}{n_0 L}\sum_{i\,\rm odd}\sum_{j\,\rm even} b_i b_j \ket{n_0 n_0 \ldots},\\
\ket{\psi_{\mathrm{Parts}}}&=&\frac{2}{(n_0+1) L}\sum_{i\,\rm odd}\sum_{j\,\rm even} b_i^{\ct} b_j^{\ct} \ket{n_0 n_0 \ldots},
\label{SF_psi}
\end{eqnarray}
where the two additional holes/particles are not spatially correlated.
We compute the energies of both states obtaining:

\begin{eqnarray}
E_{\mathrm{Holes}}&=&2\frac{\mu}{U}-2(n_0-1)+L \left[\frac{1}{2} n_0 (n_0-1)-\frac{\mu}{U} n_0\right]\nonumber\\
&&-\frac{4t}{U} n_0 \left(1-\frac{X}{t}n_0\right)\nonumber\\
&&-\frac{4t^2}{U^2}n_0(n_0+1)\left[1-\frac{X}{t} (n_0-1)\right]^2\nonumber\\
&&+\frac{2t^2}{U^2}(n_0+1)(3n_0+1)\left(1-\frac{X}{t} n_0\right)^2\nonumber\\
&&-\frac{2t^2}{U^2}L n_0(n_0+1)\left(1-\frac{X}{t} n_0\right)^2,
\label{SF_EH}
\end{eqnarray}
for to uncorrelated holes, and
\begin{eqnarray}
E_{\mathrm{Parts}}&=&-2\frac{\mu}{U}+2 n_0+L \left[\frac{1}{2} n_0 (n_0-1)-\frac{\mu}{U} n_0\right]\nonumber\\
&&-\frac{4t}{U} (n_0+1) \left(1-\frac{X}{t}n_0\right)\nonumber\\
&&-\frac{4t^2}{U^2}n_0 (n_0+1) \left[1-\frac{X}{t} (n_0+1)\right]^2\nonumber\\
&&+\frac{2 t^2}{U^2}n_0(3 n_0+2)\left(1-\frac{X}{t} n_0\right)^2\nonumber\\
&&-\frac{2t^2}{U^2}L n_0(n_0+1)\left(1-\frac{X}{t} n_0\right)^2
\label{SF_EP}
\end{eqnarray}
for two uncorrelated particles.
In the system under consideration another possible variation from the Mott state is a superfluid with two paired holes/particles. To check which one, the standard or the paired-holes/particles superfluid, is energetically more favorable we take the following states:
\begin{eqnarray}
\ket{\psi_{\mathrm{PHoles}}}&=&\frac{1}{n_0\sqrt{L}}\sum_i a_i a_{i+1} \ket{n_0 n_0 \ldots},\\
\ket{\psi_{\mathrm{PParts}}}&=&\frac{1}{(n_0+1)\sqrt{L}}\sum_i a_i^{\dagger} a_{i+1}^{\dagger} (-1)^{i} \ket{n_0 n_0 \ldots}.
\label{PSF_psi}
\end{eqnarray}
and again compute the energies perturbatively up to the second order in $t/U$.
\begin{eqnarray}
E_{\mathrm{PHoles}}&=&2\frac{\mu}{U}-2(n_0-1)+L \left[\frac{1}{2} n_0 (n_0-1)-\frac{\mu}{U}n_0\right]\nonumber\\
&&-\frac{2t}{U} n_0 \left[1-\frac{X}{t}(n_0-1)\right]\nonumber\\
&&+\frac{t^2}{U^2}(n_0+1)(7 n_0+1)\left(1-\frac{X}{t} n_0\right)^2\nonumber\\
&&-\frac{t^2}{U^2}(n_0+1)(n_0-1)\left[1-\frac{X}{t} (n_0-1)\right]^2\nonumber\\
&&-\frac{2t^2}{U^2}L n_0(1+n_0) \left(1-\frac{X}{t} n_0\right)^2,
\end{eqnarray}
\begin{eqnarray}
E_{\mathrm{PParts}}&=&-2\frac{\mu}{U}+2 n_0+L \left[\frac{1}{2} n_0 (n_0-1)-\frac{\mu}{U} n_0\right]\nonumber\\
&&+\frac{2t}{U}(n_0+1)\left[1-\frac{X}{t}(n_0+1)\right]\nonumber\\
&&+\frac{t^2}{U^2}n_0(7 n_0+6)\left(1-\frac{X}{t}n_0\right)^2\nonumber\\
&&-\frac{t^2}{U^2}n_0(n_0+2)\left[1-\frac{X}{t}(n_0+1)\right]^2\nonumber\\
&&-\frac{2t^2}{U^2}L n_0(1+n_0) \left(1-\frac{X}{t} n_0\right)^2.
\label{PSF_E}
\end{eqnarray}
In Fig.\ref{PhaseDiags} on top of the phase diagrams obtained by the cluster Gutwiller method, which we will discuss in the next section, we show the boundaries between the Mott insulator and the two possible types of the superfluid obtained by putting to zero the energy difference $E_{\mathrm{Parts/Holes}}-E_{\mathrm{Mott}}$ and $E_{\mathrm{PParts/PHoles}}-E_{\mathrm{Mott}}$. 

Apart from the already discussed PHSF and PPSF, the system also supports the phase-separated superfluid phase, {\it i.e.} such that the particles belonging to each sublattice concentrate in spatially disjoint regions and form two separate superfluids on lattices of length $L/2$. Here we estimate the value of $t/U$ for which such phase is energetically more favorable than the standard superfluid at fixed total density $\rho=N/L$. Assuming that in the superfluid phase with weak interaction only the zero momentum mode is occupied, the kinetic energy is $-2t/U \rho L$. Further, since only half of each sublattice is occupied, the average number of particles per site on occupied sites is $2\rho$, {\it i.e.} $n_0=\rho L$ distributed equally over $L/2$ sites ($L/4$ occupied sites on each sublattice). We obtain the following estimate for the energy of the superfluids cramped up on a half of each sublattice
\begin{eqnarray}
E_{\mathrm{sep}}&=&-\frac{2t}{U} \rho L + \frac{1}{2} (2 \rho)(2 \rho-1) \frac{L}{2} - \frac{\mu}{U} \rho L \nonumber\\
      &=& \left[-\frac{2t}{U}\rho + \frac{1}{2} \rho (2 \rho-1)-\frac{\mu}{U} \rho\right] L.
\end{eqnarray}

To estimate the energy of the standard superfluid in the presence of the correlated hopping, we assume to be in a regime of parameters in which the correlated hopping only modifies the tunneling rate but does not change the nature of the superfluid, as shown by the exact diagonalization data. We obtain the following
\begin{eqnarray}
E_{\mathrm{SF}}&=&-\frac{2(t-X \rho)}{U} \rho L  + \frac{1}{2} \rho(\rho-1) L - \frac{\mu}{U} \rho L \nonumber\\
      &=& \left[-\frac{2(t-X \rho)}{U} \rho + \frac{1}{2} \rho(\rho-1) - \frac{\mu}{U} \rho\right] L.
\end{eqnarray}
The system is in the standard superfluid phase whenever $E_{\mathrm{SF}}<E_{\mathrm{sep}}$, which leads to a simple criterion for the extension of the standard superfluid
\begin{equation}
    U>4X.
    \label{SFextension}
\end{equation}


\subsection{Cluster mean-field approach}\label{sub_phasediags}

We now turn to our main results, \textit{i.e.} the full phase diagram obtained within a cluster mean-field (CM) approach. The  mean-field decoupling of the single particle hopping yields a sufficient description of the Mott-insulator to superfluid transition in the standard BHM. However, given the additional correlations induced by occupation number dependent hopping, we need to use the multi-site extention (CM) of the standard Gutzwiller ansatz for a more accurate decoupling and inclusion of quantum fluctuation effects. 

The variational CM ansatz is a product state $\ket{\Psi_G}=\prod \ket{\psi}$, where each $\ket{\psi}$ is a linear combination of $d$-site Fock states $\{\ket{\mathbf n}\}=\{\ket{n_1,\ldots,n_d}\}$ ($n_i=1,2,\ldots,n_{\rm max})$
\begin{equation}
\ket{\psi}=\sum_{\mathbf{n}} f_{\mathbf{n}}\ket{\mathbf{n}}.
\label{ansatz}
\end{equation}
The Hamiltonian for the $d$-site cluster is the same as (\ref{eq_model}) in the case of on-site interaction terms and in the interior of the cluster $\mathcal{S}=\{3,\ldots,d-2\}$. The boundary sites are coupled to the rest of the system via mean-field parameters $\av{b_0},\av{b_{d+1}},\av{b_{-1}n_0},\av{n_{d+1}b_{d+2}}$:
\begin{eqnarray}
H_{\mathbf{mn}}^{\rm cluster}&=&\braket{\mathbf{m}\Bigg |H_{\mathcal{S}}+\frac{U}{2}\sum_{i=1}^{d} n_i (n_{i}-1)\Bigg|\mathbf{n}} \label{Hcluster}\\
&-&t \braket{\mathbf{m}\Bigg |\left[\av{b_0^{\dagger}}\left(1-\frac{X}{t}n_{1}\right)b_2+h.c.
\right] \Bigg|\mathbf{n}}\nonumber\\
&-&t\braket{\mathbf{m}\Bigg |\left[b_{d-1}^{\dagger}\left(1-\frac{X}{t}n_{d}\right)\av{b_{d+1}}+h.c.
\right] \Bigg|\mathbf{n}}\nonumber\\
&-&t\braket{\mathbf{m}\Bigg |\left[\av{b_{-1}^{\dagger}\left(1-\frac{X}{t}n_{0}\right)}b_1+h.c.
\right]\Bigg|\mathbf{n}}\nonumber\\
&-&t\braket{\mathbf{m}\Bigg|\left[b_{d}^{\dagger}\av{\left(1-\frac{X}{t}n_{d+1}\right)b_{d+2}}+h.c.
\right] \Bigg|\mathbf{n}}.\nonumber
\end{eqnarray}
The $d$-site ground state wave function of (\ref{Hcluster}) is computed self-consistently. First, a generic initial state of the cluster with $f_{\mathbf{n}}^{0}=1/|\{\ket{\mathbf{n}}\}| $ is used to compute the mean-field parameters. Subsequently, a new ground state is obtained and the mean-field parameters are updated. The convergence criterion to be fulfilled is $\sum_{\mathbf{n}}\left|f_{\mathbf{n}}^{i}-f_{\mathbf{n}}^{i-1}\right|<10^{-7}$, where $i$ is an index of the iteration.
From the obtained wave vector we compute the average quantities $\av{n_i},\av{n_i^2},\av{b_i},\av{b_i^2}$ on the most central sites, as being minimally coupled to the mean-field they should most accurately describe the properties of the system. However, we also examine the cluster as a whole in  search of any spatial patterns and correlations of the local observables.

The results obtained by CM are depicted in Fig. \ref{PhaseDiags} for a cluster of size $d=6$ and maximal on site occupation $5$. Additionally, for the purpose of efficient self-consistent computation, we loose the constraint of equal occupation of sublattices and allow the number of particles in the even/odd lattice to differ by one. The results show, however, that only the vectors with equal number of particles in both sublattices contribute to the final wave function describing the cluster. The main information is extracted from contour plots of the superfluid parameter combined with properties of the wave function of the cluster. Additionally, in Fig. \ref{PhaseDiags} we plot the boundaries of phases resulting from the strong coupling expansion (SCE) shown in the previous section. As visible, the CM method reveals the appearance of a rich phase diagram for different values of $X/t$. At already intermediate ECH values, $0.5<X/t\leq 0.75$, the phase diagram in Fig. \ref{PhaseDiags} b) shows a totally different scenario with respect to the $X=0$ in Fig. \ref{PhaseDiags} a). Indeed it is possible to notice that, for such ECHs, the shape of the Mott lobe is modified and a phase separated state appears for large on-site repulsion. For larger values of $X/t$ we observe that the MI is surrounded ny PHSF and PPSF. Finally in the limit of large ECHs $X/t \geq 1.05$, as confirmed also by ED calculations, the Mott lobe remains surrounded by the PHSF and PPSF but in addition to this there is also a direct transition to the droplet superfluid phase. For $X/t>0.5$ and large $t/U$, where we expect to find the droplet superfluid phase, we observe diverging particle number. It is suggestive of the droplet phase, however we were not able to obtain conclusive results. 


\section{Concluding remarks}

In this paper, we have studied a one dimensional version of the Bose Hubbard model with extended correlated hopping. By combining both numerical and analytical approaches, we have been able to show the intriguing effect played by this kind of non-local interaction. In particular the study of clustering quantities, obtained combining different and complementary approaches, turned out to be particularly meaningful. Indeed we have been able to reveal the presence of non-trivial clustered gapless phases and hence allowing us find non-standard types of the superfluids like the paired-holes and paired particles superfluid as well as the unbounded droplet superfluid. We finally stress that, as already specified, nowadays clustering properties of ultracold bosonic systems are a particularly timely topic since they are associated to non-standard state of matter \cite{pfau,ferlaino1,leticia,fattori,tanzi,chomaz,pfau1}. For this reason it is worth to underline that the discovery of novel interaction terms, like extended correlated hopping processes, which can give rise to intriguing clustered superfluid states can shed some further light in this timely and fascinating research field.


\begin{acknowledgements}
We thank M. Gajda and T. Sowi\'nski for discussion and valuable comments. J. S. acknowledges support from the National Science Center, Poland Grant No. 2015/16/S/ST2/00445. R.W.C. acknowledges support from the Polish National Science Centre (NCN) under Maestro Grant No. DEC2019/34/A/ST2/00081. M. L. and L. B. acknowledge support from ERC AdG NOQIA, Spanish Ministry of Economy and Competitiveness ("Severo Ochoa" program for Centres of Excellence in R\&D (CEX2019-000910-S), Plan National FISICATEAMO and FIDEUA PID2019-106901GB-I00/10.13039 / 501100011033, FPI), Fundaci\'o Privada Cellex, Fundaci\'o Mir-Puig, and from Generalitat de Catalunya (AGAUR Grant No. 2017 SGR 1341, CERCA program, QuantumCAT U16-011424, co-funded by ERDF Operational Program of Catalonia 2014-2020), MINECO-EU QUANTERA MAQS (funded by State Research Agency (AEI) PCI2019-111828-2 / 10.13039/501100011033), EU Horizon 2020 FET-OPEN OPTO- Logic (Grant No 899794), and the National Science Centre, Poland-Symfonia Grant No. 2016/20/W/ST4/00314.
\end{acknowledgements}



\end{document}